# COPER: a Query-adaptable Semantics-based Search Engine for Persian COVID-19 Articles

Reza Khanmohammadi*
Computer Engineering Department
University of Guilan
Rasht, Iran
rezanecessary@gmail.com

Mitra Sadat Mirshafiee
Industrial Engineering Department
Alzahra University
Tehran, Iran
mitra.mirshafiee@gmail.com

Mehdi Allahyari
Data Science Department
Wells Fargo
Atlanta, USA
mehdi.allahyari@wellsfargo.com

*Abstract*— With the surge of pretrained language models, a new pathway has been opened to incorporate Persian text contextual information. Meanwhile, as many other countries, including Iran, are fighting against COVID-19, a plethora of COVID-19 related articles has been published in Iranian Healthcare magazines to better inform the public of the situation. However, finding answers in this sheer volume of information is an extremely difficult task. In this paper, we collected a large dataset of these articles, leveraged different BERT variations as well as other keyword models such as BM25 and TF-IDF, and created a search engine to sift through these documents and rank them, given a user's query. Our final search engine consists of a ranker and a re-ranker, which adapts itself to the query. We fine-tune our models using Semantic Textual Similarity and evaluate them with standard task metrics. Our final method outperforms the rest by a considerable margin.

## I. INTRODUCTION

During the first semester of 2020 in which Severe Acute Respiratory Syndrome Coronavirus 2 (SARS-CoV-2) outbroke globally, more than 23,000 COVID-19-related research articles were published [1]. Meanwhile, as the spread of the COVID-19 passed a momentous milestone, informing the public became an international health emergency and to keep our knowledge of the existing conditions posted with the latest updates, became a high priority on the public's agenda. This trend put a spotlight on online healthcare magazines. They are constantly trying to inform people of symptoms, diagnosis, and treatments of COVID-19 and mitigating the demand for hospital services. However, due to the large volume of articles, one may go through a plethora of articles to find a piece of specific information. This highlights the significance of efficient search methods and the different advantages that they offer.

Alongside China, South Korea, and Italy, Iran was struck during the pandemic's first wave as well [2]. In the meantime, online Iranian healthcare magazines played a pivotal role in the public's welfare by dedicating themselves to inform people of the virus in terms of prevention, medical care, and applied restrictions. In this work, we collected a corpus of 3,500 articles from such magazines, developed a search engine, and established a firm baseline for **Per**sian **CO**VID-19 article retrieval called **COPER**. Our approach consists of an inexpensive ranker (BM25) and a re-ranker which utilizes a keyword and a semantic model. Particularly, we combine TF-IDF with a variation of BERT [3] to benefit from both term frequencies and contextualized information of articles. Then we calculate a *joint similarity score* for each article, which outputs a new score by taking in both TF-IDF and BERT similarities and weighing them based on the user query syntax.

Pretrained contextualized language models' robustness in capturing the semantic of an input text has yielded remarkable results in Information Retrieval (IR) [4]–[7]. Nevertheless, the absence of monolingual Transformer-based language models interfered with a similar progress in the Persian Language. ParsBERT [8], the first monolingual instance of BERT for the Persian Language that has been recently introduced, paved the way to take advantage of the context and nuance of words. We found this a great opportunity and proposed a method to develop a Semantics-based search engine to improve COVID-19 article search by experimenting with different pretrained models and amalgamating them with a keyword model (namely, TF-IDF). To semantically rank documents, not only have we experimented with ParsBERT (which we call BERT hereafter), but also we have utilized Sentence-Transformers [9] which are among the most recent achievements of Persian Natural Language Processing at the time of writing this paper.

Lack of computational resource and evaluation metrics have been among the most severe impediments to slow down research in Persian IR. To overcome the former, we proceed with the articles by overlooking their body part. Instead, we used a Keyword Extraction model to avoid processing the entire body and instead considered its keywords as its representatives. Additionally, as a secondary contribution of this paper, we developed a SICK [10] alike dataset called PerSICK and trained a classifier to measure Semantic Textual Similarity to overcome the latter. This dataset helps us evaluate ranking approaches using metrics such as Precision, Mean Average Precision, and normalized Discounted Cumulative. It consists of 3,000 sentence pairs where each pair is given a point (on a scale of 1 to 5) in terms of relatedness in meaning. We used BERT to perform a multiclass classification over PerSICK data and successfully achieved the accuracy of 98% in this task.

In brief, our contributions in this paper are as follows:

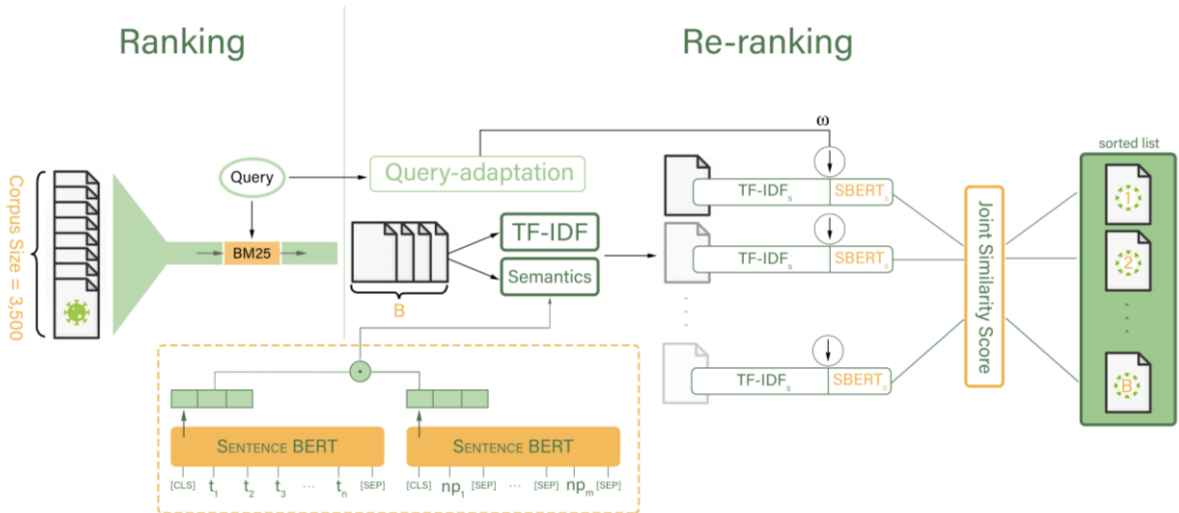

Fig. 1. The overall approach of our proposed method

- We present a new dataset by gathering COVID-19-related articles from Iranian Online Healthcare Magazines.

- We create a strong baseline for COVID-19 article search, which comprises both semantic and keyword models. We also introduce a query-adaptable search engine that combines the semantics and term frequencies of articles in its re-ranking stage as our final approach.

- We collect and represent a new dataset called PerSICK, which is, to our knowledge, the first that measures STS in Persian and helps researchers better evaluate their Natural Language Inference (NLI) tasks.

- We have made both datasets and all implementations publicly available[1], and hereby we hope this work aids both the public and the scientific community during the outbreak and thereafter.

## II. LITERATURE REVIEW

Since the outset of the pandemic, different English datasets have been introduced to tackle various COVID-19 literature-oriented tasks, among which the COVID-19 Open Research Dataset Challenge (CORD-19) [11] has been most studied. At the time of writing this paper, the dataset consists of more than 400,000 scientific papers on COVID-19 [11] and it receives daily updates from different valid resources such as PubMed and preprint servers like bioRxiv and medRxiv. CORD-19 has significantly contributed to IR, Document Search, Question Answering [12], and Text Summarization [13] systems.

In terms of IR and document search, Covidex [14] provided an openly accessible application and a firm infrastructure for researchers to analyze the CORD-19 literature in greater depth. Influenced by Multi-stage search architectures' long-lasting success [15]–[17], it incorporated different IR techniques such as BM25 and Anserini [18], [19] as its retrieval components and a fine-tuned T5 Transformer [20] on the MS-MARCO dataset [21] as its ranker. On the other hand, SLEDGE-Z [6] introduced a zero-shot baseline by utilizing BM25 and SciBERT [22] (a pretrained transformer on scientific literature) as its ranking and re-ranking components, respectively. Similar to Covidex, SLEDGE-Z fine-tunes its pretrained model on MS-MARCO due to lack of training data and applies its method on COVID-19 literature in zero-shot setting. As an efficient model in ad-hoc document retrieval, BM25 accompanies TF-IDF and Siamese-BERT [23] in the retrieving stage of a semantic search engine called CO-Search [24], which incorporates both a multi-hop question-answering module and a multi-paragraph abstractive summarizer in its ranking stage. Primarily, the engine linearly combines TF-IDF and SBERT retrieval scores to sort documents and combines it with BM25 afterward using a reciprocal ranked fusion [25]. As [26] suggests, by treating input queries as questions and documents as answers, CO-Search modulates the retrieved scores by how similar the documents are to generated answers and summaries.

Concerning the Persian literature of COVID-19, relatively less effort has been made compared to English. [2] was among the initial attempts to track the public's response to the pandemic by analyzing Persian tweets, using NLP techniques. Subsequently, [27] generally demonstrated how significant Iranian researchers' publications have been during the pandemic by searching through reliable databases such as Web of Science, Scopus, and PubMed, which depicts its large work scope. It is noteworthy that, to the best of our knowledge, we are the first to tackle document search on Persian COVID-19 literature and to make use of Transformers to employ the role of semantics in ranking documents.

Motivated by the success of the aforementioned datasets, utilized pretrained models and multi-stage search architectures, we decided to collect a dataset and establish a strong baseline for IR and document ranking approaches in Persian COVID-19 literature, which is yet proper to be contributed to.

---

[1] https://github.com/Ledengary/COPER

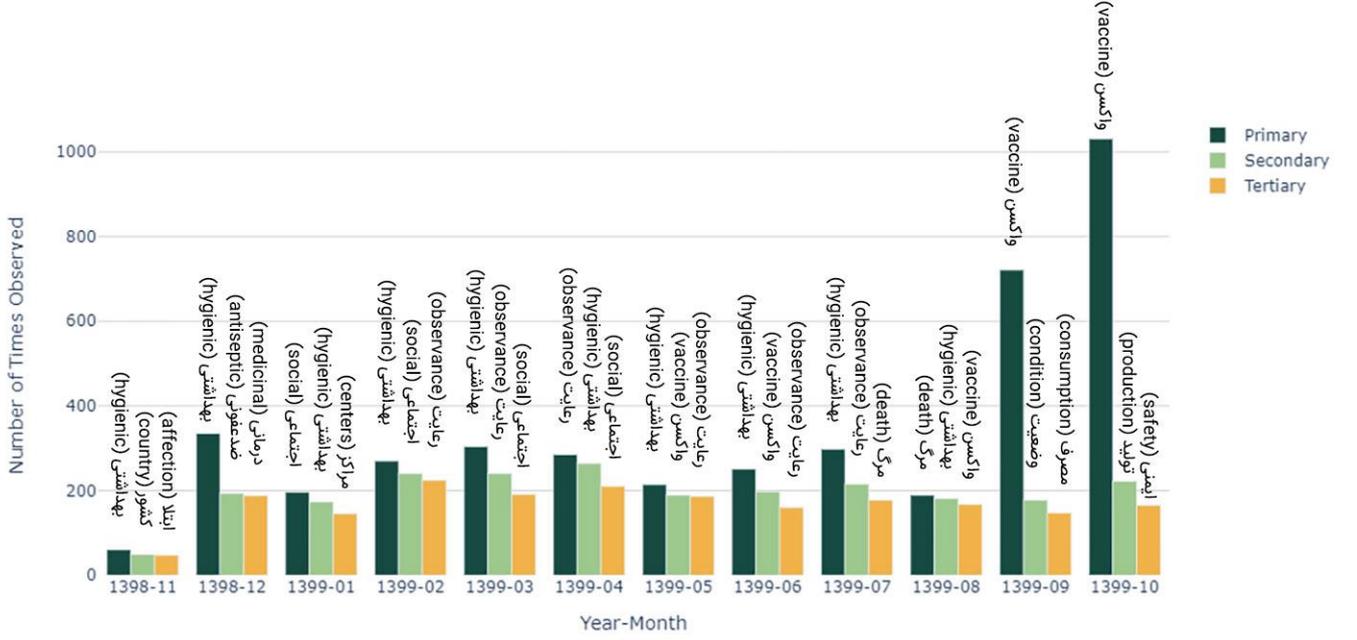

Fig. 2. Top 3 most frequently used words among extracted noun phrases during the course of a year

### III. COPER

We collected COVID-19 articles from hidoctor[2] and salamatnews[3], two of the most nationally known online healthcare magazines. Our dataset consists of 3,500 articles, covering a wide range of topics (e.g., symptoms, treatments, prevention, general related knowledge, and national updates). First, we applied standard preprocessing filters such as stopwords removal and space correction[4], we then index dataset articles using Faiss [28], a library for efficient similarity search and dense vectors clustering. Building upon multi-stage document ranking approaches' success, we introduce COPER, a framework consisting of a ranking and a re-ranking stage to respectively benefit from collecting a high-recall set of candidates and a high-precision set of final results. An overview of our approach is depicted in Fig. 1.

#### A. Ranker

We used BM25 to retrieve documents. Given a query *Q*, which contains query terms *q*, this scoring function calculates a linear weighted combination of each query term's score and outputs the sum of their scores. The BM25 score of document *D* and query *Q* is obtained through calculating Equation (1).

$$score(D, Q) = \sum_{i=1}^{n} IDF(q_i) \cdot \frac{f(q_i, D) \cdot (k_1 + 1)}{f(q_i, D) + k_1 \cdot \left(1 - b = b \cdot \frac{|D|}{avgdl}\right)} \quad (1)$$

In the above equation, $f(q_i, D)$ and $IDF(q_i)$ stand for $q_i$'s Term and Inverse Document Frequencies, which will be discussed in the following subsection. In addition, $|D|$ and $avgdl$ as document *D*'s number of words and average document length of our corpus defines how long document *D* is relative to our corpus's average document length. Moreover, $k_1$ and $b$ are free parameters that have experimentally shown us to be most productive by being set to 1.5 and 0.75. Since related items are mostly aggregated among the top 800 documents of BM25's generated list of candidate documents, we set a threshold to decrease the number of articles that need to be re-ranked. Our experiments demonstrated that very few related items are likely to be found beyond this threshold.

#### B. Re-ranker

Our primary goal here is to employ the semantics when re-ranking the candidates' list. To accomplish that, we amalgamate TF-IDF, a keyword ranking model, with Sentence BERT as our semantics model. Correspondingly, we avail the approach of both articles' context and term frequencies.

TF-IDF is used as a weighting factor for textual features and is structured to evaluate how relevant a term is to a particular document among a certain set of documents. The TF-IDF vector (Equation 4) is obtained by multiplying the Term Frequency (*TF*) (Equation 2) and the Inverse Document Frequency (*IDF*) (Equation 3).

---

[2] https://www.hidoctor.ir
[3] https://www.salamatnews.com
[4] Adjusting Non-Joint Zero Width (NJZW) letters

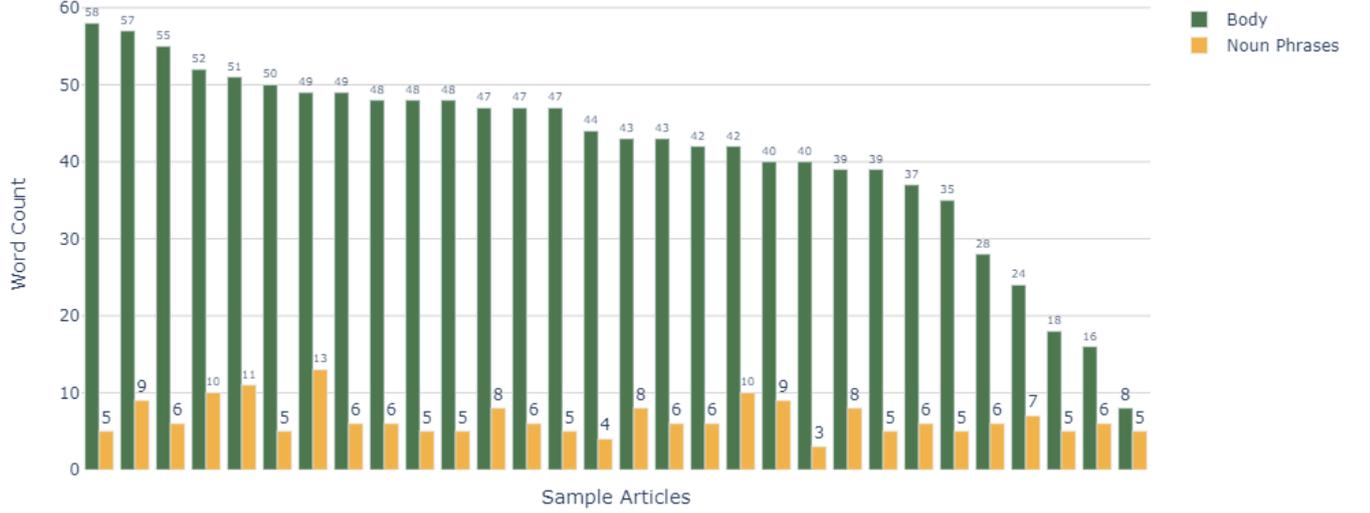

Fig. 3. A comparison between the word counts of 30 samples' body and extracted noun phrases

$$TF(t, D) = \log(1 + freq(t,d)) \quad (2)$$

$$IDF(t, D) = \log(\frac{N}{count(d \in D : t \in d)}) \quad (3)$$

$$TF - IDF = TF(t,d) \cdot IDF(t,D) \quad (4)$$

Given the number of times term t appears in document d as *freq(t,d)*, we calculate term frequency. Also, we use equation (3) to measure how rare *t* is in our corpus (*D*),. Therefore, we have access to textual features that may guide the semantics model to better rank candidates.

Even though we are working on a closed domain problem, our dataset articles' essence follows a generic tone, i.e. there is no need for additional medical knowledge. This facilitates document ranking since no pretrained monolingual language models on Persian medical or scientific literature are available at the time of writing to capture the domain-specific semantics. Following Vanilla BERT's success in employing contextual information of text to rank documents [7], we leverage Sentence BERT (SBERT) to embed our articles contextually.

Unlike BM25 and TF-IDF, SBERT is a computationally expensive model with a specific token limitation of 512. We chose to proceed with an article's keywords instead of its entire body text, to overcome the challenge of processing the whole document and the lack of computational resources. We used Yake! [29], an unsupervised multilingual keyword extractor, which generates a sequence of candidate keywords by assigning a score *S(kw)* to each. Given *S(w)* as each term's score, we obtain a phrase's final score by calculating Equation (5).

$$S(kw) = \frac{\prod_{w \in kw} S(w)}{TF(kw) \times (1 + \sum_{w \in kw} S(w))} \quad (5)$$

Extracting high-quality keywords relies heavily on text cleaning. Using BERT, we remove specific named entities such as *place* or *person* to guide our keyword extraction model better.

However, to infer semantics from keywords better, we output the entire original noun phrase that a keyword has been embedded in and not just the keyword itself. This approach has shown to be more capable of conceiving hidden contextual information, which yields better overall ranking results. Fig. 2 and Fig. 3 are illustrations that highlight the applicability of this technique in identifying frequent words and lightning the computational load.

BERT is pretrained using two predefined tokens of *[SEP]* and *[CLS]*, enabling the model to segment the input text and make predictions, respectively. By segmenting, BERT can simultaneously judge different text pieces and provide us a final representation using the *[CLS]* token. Having extracted keywords' related noun phrases (*NP*), we then separately pass an article's title (*T*), containing $t_{1-n}$ tokens, and *NP*, containing $np_{1-m}$ phrases to SBERT to achieve contextual embeddings of both title and body. We observe that an article's title plays more effectively than noun phrases when ranking. Hence, we experimented with different values and realized that multiplying the title's vector by 1.1 and the noun phrases' vector by 1 output the most acceptable results. Lastly, we concatenate the two vectors to achieve a final semantics-based representation.

When a user enters a query and the ranker defines a set of candidate documents, we calculate a semantics and TF-IDF vector for each candidate in the re-ranking stage. To measure the similarity between a query's vector and the candidates' TF-IDF/semantics vector, we use the *cosine similarity* metric to measure the angle between the two. We named TF-IDF and semantics vectors' similarity scores as *TF-IDF$_s$* and *SBERT$_s$*. We additionally defined a *join similarity score* (*jss*) (Equation 6) as the final step to re-rank candidates, which takes the two *TF-IDF$_s$* and *SBERT$_s$* scores as inputs and generates a *jss* score as output. Lastly, the documents are sorted by their *jss* values and are returned as the search engine's final output.

$$\begin{aligned} jss(\omega, TF - IDF_s, SBERT_s) \quad & (6) \\ = \omega \cdot (TF - IDF_s) & \\ + (1 - \omega) \cdot (SBERT_s) & \end{aligned}$$

To make the system adaptable to a user's query, we defined a *wordiness level* ($\omega$) to weight the two *TF-IDF$_s$* and *SBERT$_s$* scores differently. The intuition here is that a user query is not necessarily a meaningful piece of text from which SBERT is likely to extract high-quality contextual information. Hence, we defined $\omega$ to decrease/increase the semantics/TF-IDF vectors' effect in such scenarios (Fig. 4). We calculate $\omega$ using part-of-speech tags (POS) and their specific patterns, introduced in [30] to extract semantic information from Persian text. We successfully adapted their proposed method of Relation Extraction (RE) to generate triples using Regex. We then heuristically calculate $\omega$, which is the ratio of wordiness in a user's query.

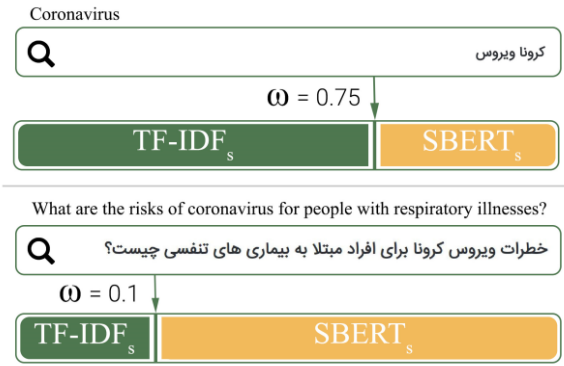

Fig. 4. Samples of Query-adaptation with $\omega$ value

In further steps we experiment with both keyword and semantic models and explore each of their efficiencies and demonstrate how our proposed multi-stage ranking approach outperforms the others on our dataset. Official task metrics such as Precision and normalized Discounted Cumulative Gain mostly demand text relevance as a pre-requisite. However, a dataset with a pre-defined set of queries and their respective group of relevant/not relevant documents is not currently available in our domain of study. Hence, we decided to measure the similarity of ranked documents with any query using Semantic Textual Similarity (STS) to classify any ranked document as either relevant or not relevant.

### C. PerSICK

As one of the most crucial components of Information Retrieval, STS provides a solution on how to measure the similarity between text. Among introduced STS datasets [31]–[36], the Sentence Involving Compositional Knowledge (SICK) dataset [9] proposed a standard scale of 1 (totally irrelevant) to 5 (complete semantic equivalence) to score the similarity between text pairs (readers are referred to [10] for more detail about the scale of ranking). We translated a random subset of 3,000 out of SICK's 10,000 sentence pairs and subsequently trained a multi-class classifier to categorize a given text pair from 1 to 5 semantically. We utilized BERT to let its contextualized word embedding take care of finding matches and mismatches between two sentences' semantics. This model, which we refer to as the **ESBERT** (**E**valuator **S**entence **B**ERT) model hereafter, successfully achieves 97.25% accuracy in classifying text pairs. Both train (2,000 pairs) and test (5,000 pairs) sets are available in the project repository with which future work may continue experimenting with as a shared task evaluator.

### IV. EXPERIMENTS

#### A. Datasets

Overall, we use three different datasets, all of which are broken down in Table I.

TABLE I. UTILIZED DATASETS IN OUR RESEARCH

| Dataset | Use cases |
| --- | --- |
| COVID-19 articles | Document Search |
| PerSICK | Training ESBERT, fine-tuning Bert-based models |
| WHO | Evaluating models |

It is noteworthy that 2,500 sentence pairs were used to train and test **ESBERT** and the remaining 500 as semantically similar sentence pairs to fine-tune BERT-based models. Also, we assess our models using 50 Persian Frequently Asked Questions (FAQs) on World Health Organization's (WHO) website[5] in order to avoid any biases and fully demonstrate each model's efficiency in a real-world scenario. The queries cover a wide range of COVID-19-related issues and wordiness levels, making it best to highlight each model's advantages and disadvantages.

#### B. Evaluation Metrics

In an analogy to similar English COVID-19 datasets, ours consists of relatively fewer data, and the number of ground-truth-relevant articles may not be large for every query. Hence, we considered @10 to be a reasonable value as the cutoff of our metrics. Passing the WHO set to each model, all models are compared by the following metrics:

- **Mean Precision (MP@10):** Precision is the relevant fraction of retrieved documents whenever a query is given. We calculate all precisions' mean to demonstrate how successful a model has been in retrieving relevant articles.

- **Mean Average Precision (MAP@10):** We use MAP to capture the relevancy flow among retrieved articles.

- **normalized Discount Cumulative Gain (nDCG@10):** Unlike other metrics, nDCG is not limited to a relevant (1) and not relevant (0) scale and supports other scales of relevancy and assumes relevant documents are valuable as they appear at the top.

---

[5] https://www.who.int/ar/emergencies/diseases/novel-coronavirus-2019-farsi

TABLE II. RANKING MODELS' PERFORMANCE

| Subtype | Model | ESBERT | | | | Human Assessors | | | |
|---|---|---|---|---|---|---|---|---|---|
| | | *MP@10* | *MAP@10* | *nDCG@10* | *ASTS* | *MP@10* | *MAP@10* | *nDCG@10* | *ASTS* |
| Keyword Models | BM25 | 0.5240 | 0.6872 | 0.9117 | 3.3156 | 0.5714 | 0.7182 | 0.8421 | 2.7278 |
| | TF-IDF | 0.6000 | 0.7551 | 0.9232 | 3.4287 | 0.4580 | 0.6649 | 0.8080 | 2.4308 |
| Semantic Models | BERT | 0.3300 | 0.4323 | 0.9007 | 2.9592 | 0.2789 | 0.4679 | 0.6088 | 1.7210 |
| | SBERT-WT | 0.6060 | 0.7020 | 0.9183 | 3.3456 | 0.2653 | 0.4972 | 0.9202 | 2.4263 |
| | SBERT-FT | 0.5600 | 0.7010 | 0.9202 | 3.2515 | 0.2448 | 0.4917 | 0.5656 | 1.5804 |
| | SBERT-WN | 0.6259 | 0.7752 | 0.9244 | 3.4055 | 0.7210 | 0.8176 | 0.9164 | 3.1065 |
| | SBERT-WN (fine-tuned) | 0.6340 | 0.7736 | 0.9221 | 3.4323 | 0.6326 | 0.8688 | 0.8944 | 3.2040 |
| Multi-stage Models | BM25+TF-IDF+SBERT-WN (fine-tuned) | **0.7340** | **0.8233** | **0.9492** | **3.5594** | **0.8503** | **0.9072** | **0.9449** | **3.5532** |

- **Average Semantic Textual Similarity (ASTS):** Which reveals how semantically similar retrieved articles are to the query on a scale of 1 to 5.

Given such metrics, we also asked human assessors to evaluate a total of 4,000 ranked articles by their relevancy to their respective queries.

*C. Results*

As shown in Table II, we categorize our models into three subtypes as below:

- **Keyword Models:** which contains BM25 ($k_1$ = 1.5, b = 0.75) and TF-IDF.
- **Semantic Models:** which contains BERT and three variations of SBERT trained on Wiki NLI (SBERT-WN), Wiki Triplet (SBERT-WT) and FarsTail (SBERT-FT).
- **Multi-stage Models:** which only contains our final approach, COPER.

Despite their insignificant differences, BM25 and TF-IDF have been shown to retrieve similar yet different results. BM25 generally distinguishes irrelevant articles from the relevant ones, whereas TF-IDF plays effectively in capturing specific keywords. Unlike keyword models, semantic models are not applied to the entire article; instead, they are fed as ([CLS] $t_{1-n}$ [SEP] $np_1$ [SEP] … [SEP] $np_m$ [SEP]). Nevertheless, calculating the cosine similarity of two contextualized word embeddings has also been shown to yield better results over keyword models and reduce the word count by a significant margin. Among semantic models, SBERT variations have mostly outperformed BERT, highlighting contextualized sentence embeddings' effectiveness in better ranking the documents. It is worth mentioning that our COVID-19 dataset consists of general articles with very similar literature of what BERT was trained on in the first place. Hence, we fine-tuned leading SBERT-WN on 500 sentence pairs to better guide the model in capturing the semantics from COVID-19 articles.

Our proposed system consists of two stages, which first inexpensively ranks the top 800 documents using BM25 and then re-ranks them using TF-IDF and SBERT-WN. The only difference is that instead of expecting BERT to encode both titles and keywords to a mutual vector, we divide the encoder into two separate encoders as ([CLS] $t_{1-n}$ [SEP]) and ([CLS] $np_1$ [SEP] … [SEP] $np_m$ [SEP]). This method successfully outperforms the others by yielding more efficient *MP@10, MAP@10, nDCG@10,* and *ASTS* results in automatic and human evaluation processes.

V. CONCLUSION

Natural languages supported by monolingual pretrained language models have considerably advanced by taking advantage of contextualized word embeddings to better capture semantics. Since their inception in Persian NLP, such models have been likely to aid us in information retrieval and document search. We found this an opportunity to develop an adaptable semantics-based search engine and tackle the ongoing threat of COVID-19. We developed a baseline and described how the most efficient results are obtained using a multi-stage approach of both keyword and semantic models rather than using either a keyword-only or semantic-only model. As Future work, we aim to better distinguish term frequencies and semantics by improving the ω value. Additionally, training and fine-tuning more models and updating the datasets with an improved computational resource may as well benefit both the public and researchers in certain means.